\documentclass[11pt]{article}
\usepackage{moriond,epsfig}

\def\NPB{{\em Nucl. Phys.} B}
\def\PLB{{\em Phys. Lett.}  B}

\def\PRD{{\em Phys. Rev.} D}

\def\be{\begin{equation}}
\def\ee{\end{equation}}
\def\bea{\begin{eqnarray}}
\def\eea{\end{eqnarray}}

\begin{document}
\vspace*{4cm}
\title{TOP PHYSICS AT LHC WITH $t\bar{t}$ EVENTS}

\author{ F. HUBAUT\\
On behalf of the ATLAS and CMS collaborations}

\address{CPPM, CNRS/IN2P3, Univ. M\'editerran\'ee, \\
163 Av. de Luminy, Case 902, 13288 Marseille Cedex 9, France}

\maketitle\abstracts{
The new CERN proton-proton collider, the LHC, is about to start in 2007 its data taking. 
Millions of top quarks will be available out of these data, allowing to perform a wide range of precision measurements and searches for new physics.
An overview of the planned top physics program accessible with $t\bar{t}$ events is given for the ATLAS and CMS experiments.
A particular emphasis is put on the precision measurements of the top mass, top polarization and searches for new physics in top production and decay.
}

%%%%%%%%%%%%%%%%%%%%%%%%%%%%%%%%%%%%%%%%%%
%%%%%%%%%%%%%% INTRODUCTION %%%%%%%%%%%%%%
%%%%%%%%%%%%%%%%%%%%%%%%%%%%%%%%%%%%%%%%%%

\section{Introduction}

The top quark plays a central role in the physics programs of present and future high energy physics collider experiments~\cite{chakra,beneke}.
It is a privileged tool for precise studies of the Standard Model (SM) of particle physics, 
being by far the heaviest fundamental particle and the only quark decaying before hadronization takes place.
As a consequence, a precise measurement of its mass constitutes a crucial test of the electro-weak sector and puts constraints onto the Higgs boson mass 
via radiative corrections, while its very short lifetime offers a unique window on bare quarks. 
Moreover, its mass, close to the electro-weak symmetry breaking (EWSB) scale, suggests that the top quark could play a special role in the EWSB mechanism and that new physics
might be preferentially coupled to it. It is therefore a sensitive probe to new physics.
Top quark events are also a major source of background for many search channels, and precise understanding of top signal is crucial to claim new physics.
Last but not least, top quark events provide an essential tool for understanding and calibrate the detector: triggering, $b$-tagging, jet energy scale can all
benefit from top signal.\\

Ten years after the top quark discovery at the Tevatron Fermilab collider, we still know quite little about its production and decay mechanisms,
because of limited statistics. This leaves plenty of room for new physics in the top quark sector.
Even if Tevatron run~II starts to incisively probe it~\cite{tevatron}, a new opportunity for top quark physics will be opened by the
CERN proton-proton facility LHC (Large Hadron Collider), 
that will collide beams with a center of mass energy of 14~TeV at a design luminosity of 10$^{34}$~cm$^{-2}$s$^{-1}$ and will start to take data in 2007.
The main reason is that the LHC will be a top factory, with cross-sections for producing $t\bar{t}$ pairs and single tops roughly
100~times larger than at Tevatron: 8 (3) millions of $t\bar{t}$ pairs (single tops) will be produced per year during its first phase, corresponding to 10~fb$^{-1}$,
the initial luminosity being about 10 times lower than the design one.
Single top, which has not been yet discovered, is described in more details elsewhere in these proceedings~\cite{single}.
We will here focus on $t\bar{t}$ channels, whose SM final states are completely determined by the $W$-boson decay: as
all top quark decay through $t \rightarrow W^+ b$, semileptonic (44\%), dileptonic (10\%) and all hadronic (46\%) $t\bar{t}$ final states are possible. 

\section{Top mass measurement}

\begin{figure}
{
\epsfig{file=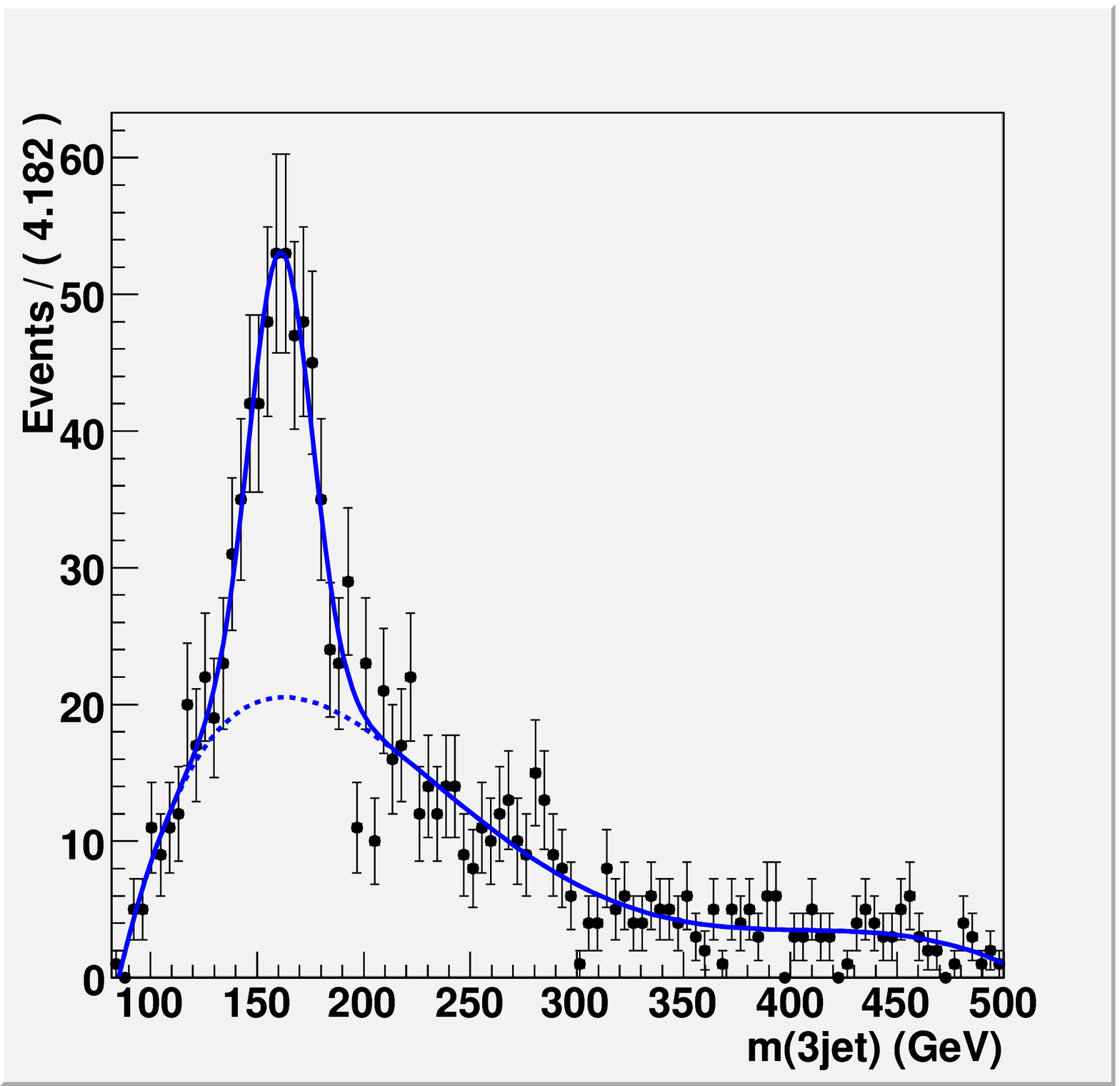, width=.45\textwidth, height=5.3cm, bbllx=0pt,bblly=0pt,bburx=530pt,bbury=500pt}
\hspace*{.6cm}\epsfig{file=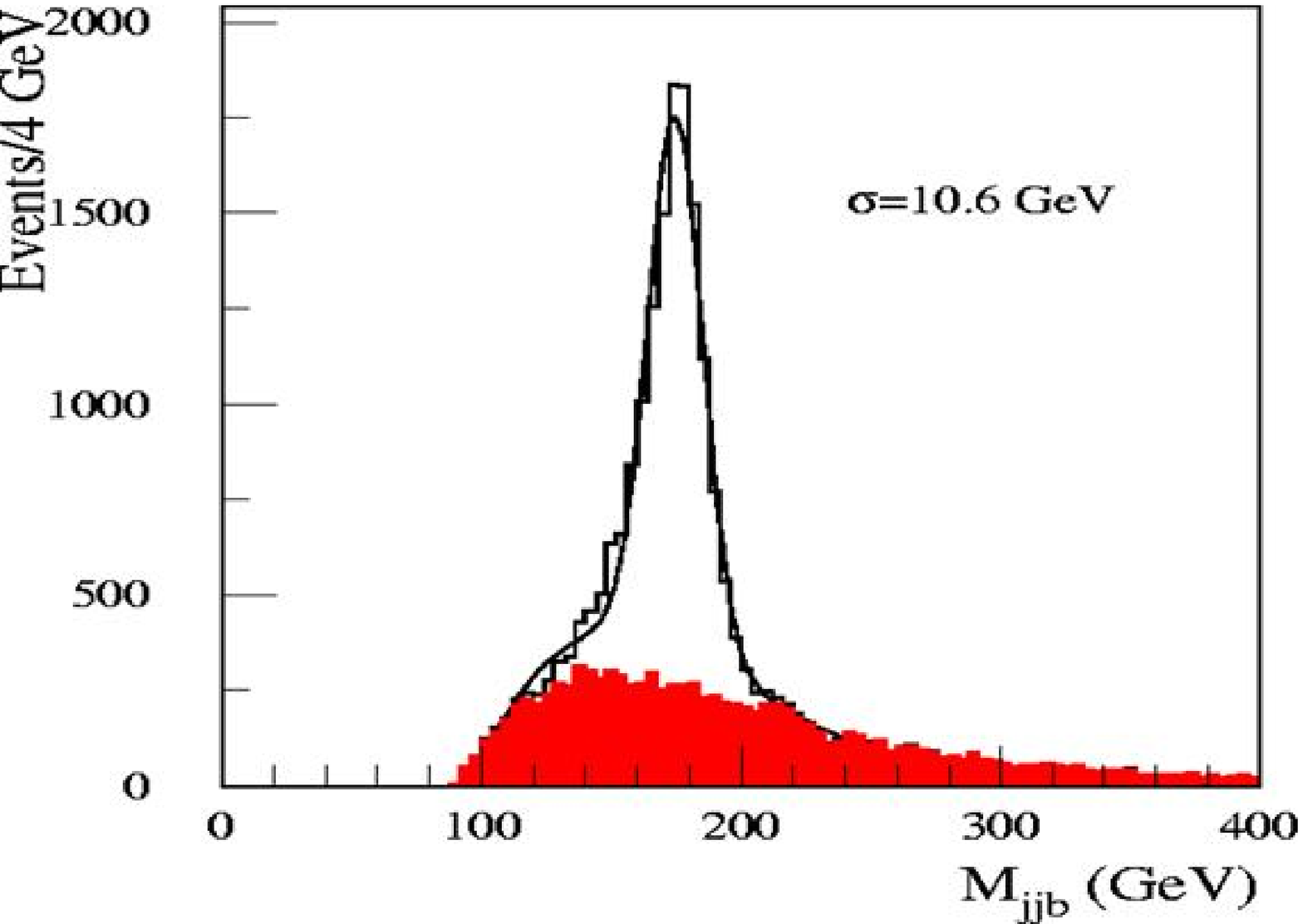, width=.49\textwidth, height=5.5cm}
\begin{picture}(1,1)
\begin{footnotesize}
\put(-370,120){$\sigma\sim15$ GeV}
\end{footnotesize}
\end{picture}
}
\caption{Reconstructed top mass with semileptonic $t\bar{t}$ events using 100~pb$^{-1}$ of simulated data assuming no $b$-tagging (left), 
and 10 fb$^{-1}$ with nominal $b$-tagging (right).
The dashed (shaded) histogram on the left (right) represents the background (combinatorics and $W$+4~jets).
}
\label{fig:topmass}
\end{figure}

The semileptonic channel provides a clean and abundant top quark sample, 2.5 million of events with electron or muon being expected with 10~fb$^{-1}$ of LHC data.
With one isolated high p$_T$ lepton to trigger on, four high p$_T$ jets, and only one neutrino, it is the most promising channel for precise measurements.
Even without $b$-tagging, a pessimistic scenario, and simply forming the 3 jets combination giving the highest p$_T$,
a clear top signal should be observed quickly above the background, dominated by the combinatorics and $W$+4~jets events. 
This is illustrated in Figure~\ref{fig:topmass} (left) obtained with 100~pb$^{-1}$ of fully simulated data, $i.e.$ one day of running at 10$^{33}$~cm$^{-2}$s$^{-1}$. 
This simple top sample extraction should enable initial measurements of cross-section and mass, 
and give important feedbacks on detector performance (such as jet energy scale and $b$-tagging) and on the reliability of Monte Carlo description.
The improvement of detector performance with data and time will increase the purity of the top sample.
This is illustrated in Figure~\ref{fig:topmass} (right), which assumes a realistic $b$-tagging with a $b$-jet efficiency of 60\% coupled 
with a light ($c$) jet rejection above 100 (10).
The full event topology is reconstructed after estimating the unknown neutrino momentum and correctly associating the 
two $b$-jets to the $b$ and $\bar{b}$ quarks.
The background is well under control, with a signal over non-$t\bar{t}$ background ratio above 50
and a dominant contribution from $t\bar{t}$ events themselves (combinatorics and $t\bar{t} \rightarrow \tau + X$ events).
With 10~fb$^{-1}$ of data, $70\,000$ events should be selected with a purity around 65\%.\\

As a consequence of such a large statistics, measurements will be dominated by systematic uncertainties.
The mass measurement is dominated by uncertainties on the $b$-jet energy scale and final state radiations. 
The latter can be reduced with a kinematic fit to the entire $t\bar{t}$ event by reconstructing both leptonic and hadronic sides. 
A total error on the top mass below 2~GeV should be feasible, possibly reaching an ultimate precision around 1~GeV.
Exploiting fully the high statistics, different sub-samples and various methods can be used 
to perform top mass measurements~\cite{TT_MASS}, leading to different source of systematic errors and allowing therefore 
reliable cross-checks. This is also the case with dileptonic and hadronic channels: the measurement is more difficult
for various reasons, but a total accuracy around 2-3~GeV seems achievable. For all present analyses, a major source of systematics
comes from the $b$-jet energy scale. This led to the development of an alternative method, based on the identification
of J/$\Psi$ originating from a $b$-quark decay and using the linear correlation between the J/$\Psi$-lepton invariant mass
and the top mass~\cite{TT_MASS,jpsi}. Due to a small branching ratio, this approach requires however a high luminosity (100~fb$^{-1}$)
to achieve a precision around 1~GeV.

\section{Top properties}

\begin{figure}
{
\epsfig{file=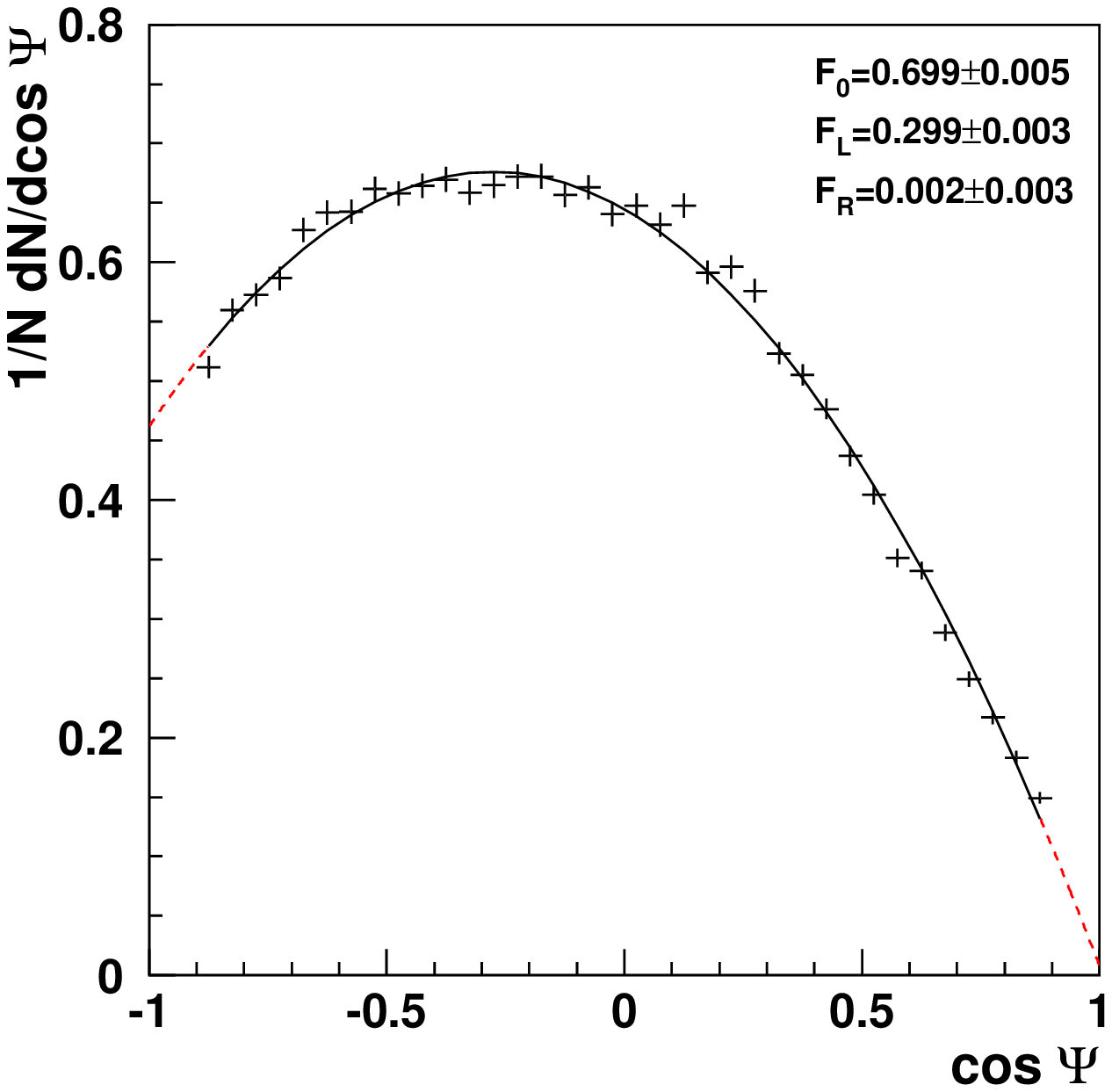, width=.53\textwidth, height=5.7cm}
\epsfig{file=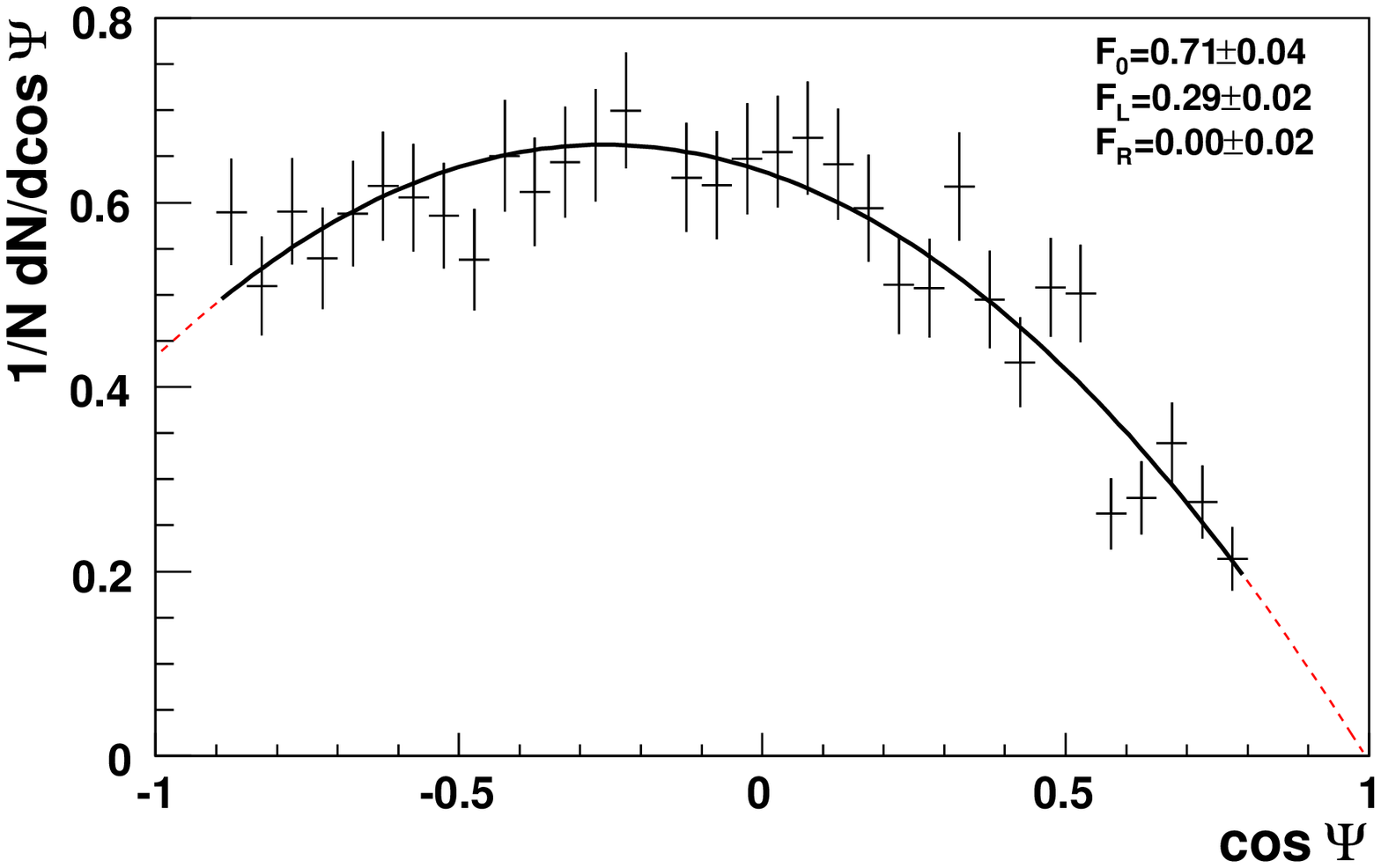, width=.49\textwidth, height=5.5cm}
}
\vspace*{-.5cm}
\caption{Reconstructed angular distribution for $W$ polarization measurement with semileptonic $t\bar{t}$ events, including background.
Events were processed through fast (left, 10~fb$^{-1}$) and detailed (right, 0.5~fb$^{-1}$) simulation of the detector.
Full lines are the results of a fit allowing to extract $W$ helicity state probabilities $F_0$, $F_L$ and $F_R$.}
\label{fig:wpola}
\end{figure}

The high statistics available at LHC will allow to study many properties of the top quark, besides its mass.
Using the same fully reconstructed $t\bar{t}$ events, it will be possible to 
%perform precision measurements allowing to 
search indirectly for new physics in top production and decay mechanisms. 
As an example, the top quark spin properties, through $W$ polarization and $t\bar{t}$ spin correlation measurements, can lead
to a deep insight of the nature of the top quark couplings to fermions and to the mechanisms (SM or not) responsible for its production.
The measurement of the SM predicted, but never measured, top spin 
correlation\footnote{At LHC, $t$ and $\bar{t}$ spins are not polarized in $t\bar{t}$ pairs, 
but rather correlated.}, with asymmetries between spin-like and spin-unlike pairs~\cite{Bernreuther}, constitutes a test of the top production.
The $W$ boson polarization measurement complements this study by testing only the top decay.
Both measurements are directly accessible through angular decay distributions in top and $W$ rest frames.
Because of its high mass, the top quark decays before hadronization or spin flip, 
thus leaving an imprint of its spin on its daughter angular distributions.
This is illustrated in Figure~\ref{fig:wpola}, showing the reconstructed $\cos\Psi$ distribution, 
where $\Psi$ is the angle between the charged lepton direction in the $W$ rest frame and the $W$ direction in the top quark rest frame.
The longitudinal ($F_0$), left ($F_L$) and right ($F_R$) $W$ helicity state probabilities are 
extracted from a two parameter fit on this distribution~\cite{Kane}.
Results from  fast (left) and detailed (right) simulated data samples are in good agreement. 
Similar angular distributions are used for spin correlation measurement~\cite{Bernreuther}.
A precision of 1-2\% is obtained for $F_0$ and $F_R$ and 4\% for the top spin correlation asymmetries 
after 10~fb$^{-1}$, as shown in Table~\ref{tab:wpola_spinco}~\cite{TT_WPOLA_SPINCO}. These results are dominated by the same systematics as 
for the top mass measurement. 
A deviation from SM values could point to anomalous $gt\bar{t}$ or $tWb$ couplings, a top spin different from 1/2, or an anomalous decay such as
$t\rightarrow H^+ b$.

\begin{table} 
\caption{ATLAS sensitivity to $W$ polarization components ($F_0$, $F_L$, $F_R$) and top spin correlation 
asymmetries ($A$ and $A_D$) after 10~fb$^{-1}$ and combination of dileptonic and semileptonic 
channels.} \label{tab:wpola_spinco} 
\vspace{0.4cm}
\begin{center}
\begin{tabular}{|l||c|c|c||c|c|} 
\hline
Parameters             & $F_0$ & $F_L$ & $F_R$ & $A$   &  $A_D$ \\ 
\hline        
SM values              & 0.703 & 0.297 & 0.000 & 0.422 & -0.290 \\ 
\hline
 Statistical errors    & 0.004 & 0.003 & 0.003 & 0.014 & 0.008  \\ 
\hline
 Systematic errors     & 0.015 & 0.024 & 0.012 & 0.023 & 0.010  \\ 
\hline 
\end{tabular} 
\end{center}
\end{table}

\section{Direct searches for new physics}

Direct searches for new physics in top production and decay mechanisms have also been investigated.
As an example, a heavy resonance decaying to $t\bar{t}$ might enhance the cross-section and produce a peak in the 
$t\bar{t}$ invariant mass spectrum, as predicted by various models beyond the SM. 
A 1~TeV ``generic'' narrow resonance~\cite{beneke} may be discovered at LHC with 30~fb$^{-1}$ of data for a cross-section down to 800~fb.
Many models beyond the SM include a complicated EWSB sector, with implications for top decays. Examples
include the possible existence of charged Higgs bosons, or possibly large flavor changing neutral currents (FCNC) in top decays.
The top decay to a light charged Higgs boson $t \rightarrow H^+ b$ can be searched for through an excess of $t\bar{t}$ events with $\tau$-jets 
or a deficit of dilepton events. $H^+$ with mass up to 160~GeV may be observed and 170~GeV may be excluded with 30~fb$^{-1}$ of data~\cite{higgsdecay}.
In the SM, FCNC top decays are highly suppressed ($BR<10^{-13}$-$10^{-10}$). However, several extensions of the SM can lead
to very significant enhancements of these $BR$s (up to $10^{-3}$), which could be detected directly in $t\bar{t}$ events.
ATLAS and CMS studies show that the current limits on the branching ratios of these processes can be improved by a factor $10^{2}$ to $10^{3}$
already with 10~fb$^{-1}$ of data~\cite{beneke}, such a sensitivity allowing to start to probe models beyond the SM.

%%%%%%%%%%%%%%%%%%%%%%%%%%%%%%%%%%%%%%%%%%
%%%%%%%%%%%%%% CONCLUSIONS    %%%%%%%%%%%%
%%%%%%%%%%%%%%%%%%%%%%%%%%%%%%%%%%%%%%%%%%
\section{Conclusions}
\label{sec:conclu}

The LHC will open a new era of precision measurements in the top quark physics that will lead to
a thorough determination of the top quark properties, as its mass, couplings and polarization.
Besides stringent tests of the SM, those measurements also constitute powerful probes in the search for new physics
and could lead to new discoveries in the first years of the LHC low luminosity data taking period.

%\section*{Acknowledgments}
%I would like to thank the conference organizers for their hospitality and financial support.
%I also thank Pascal Pralavorio (CPPM) for constant help and support.

%%%%%%%%%%%%%%%%%%%%%%%%%%%%%%%%%%%%%%%%%%
%%%%%%%%%%%%%% BIBLIO         %%%%%%%%%%%%
%%%%%%%%%%%%%%%%%%%%%%%%%%%%%%%%%%%%%%%%%%


\section*{References}
\begin{thebibliography}{99}

\bibitem{chakra} D.~Chakraborty {\it et al.}, {\em Ann. Rev. Nucl. Part. Sci. }~{\bf 53}, 301 (2003) [hep-ph/0303092].
%W.~Wagner, {\em Rept. Prog. Phys. }~{\bf 68}, 2409 (2005) [hep-ph/0507207].\\

\bibitem{beneke} M.~Beneke {\it et al.}, CERN-TH-2000-100 (2000) [hep-ph/0003033].

\bibitem{tevatron} L.~Shabalina, D.~Whiteson, J.~Leveque and A.~Gresele, these proceedings.

\bibitem{single} M. Mohammadi Najafabadi, these proceedings.

\bibitem{TT_MASS} I.~Borjanovic {\it et al.}, {\em Eur. Phys. Jour.} C~{\bf 39} S2, 63 (2005) [hep-ex/0403021].

\bibitem{jpsi} A.~Kharchilava, \PLB~{\bf 476}, 73 (2000) [hep-ph/9912320].

\bibitem{Bernreuther} W.~Bernreuther {\it et al.}, \NPB~{\bf 690}, 81 (2004) [hep-ph/0403035].

\bibitem{Kane} G.L.~Kane {\it et al.}, \PRD~{\bf 45} 124 (1992).
 
\bibitem{TT_WPOLA_SPINCO} F.~Hubaut {\it et al.}, {\em Eur. Phys. Jour.} C~{\bf 44} S2, 13 (2005) [hep-ex/0508061].

\bibitem{higgsdecay} S.~Banerjee and M.~Maity, {\em J. Phys.} G~{\bf 28}, 2443 (2002).

\end{thebibliography}
\end{document}